\documentclass[aps,twocolumn,floatfix]{revtex4-1}

\usepackage{bm}
\usepackage{amsmath,amssymb,cases}
\usepackage{braket}
\usepackage[dvipdfmx]{color}
\usepackage[dvipdfmx]{graphicx}

\usepackage{ulem}

\begin{document}

\title{Magnetoelectric effect in band insulator--ferromagnet heterostructures}

\author{Ken N. Okada}
\author{Yasuyuki Kato}
\author{Yukitoshi Motome}

\affiliation{Department of Applied Physics, University of Tokyo, Tokyo 113-8656, Japan}

\begin{abstract}
We theoretically study magnetoelectric effects in a heterostructure of a generic band insulator and a ferromagnet.
In contrast to the kinetic magnetoelectric effect in metals, referred to as the Edelstein effect or the inverse spin galvanic effect, our mechanism relies on virtual interband transitions between the valence and conduction bands and therefore immune to disorder or impurity scattering.
By calculating electric field--induced magnetization by the linear response theory, we reveal that the magnetoelectric effect shows up without specific parameter choices. 
The magnetoelectric effect qualitatively varies by changing the direction of the magnetic moment in the ferromagnet: the response is diagonal for the out-of-plane moment, whereas it is off-diagonal for the inplane moment.
We also find out that in optical frequencies, the magnetoelectric signal can be drastically enhanced via interband resonant excitations.
Finally, we estimate the magnitude of the magnetoelectric effect for a hybrid halide perovskite semiconductor as an example of the band insulator and compare it with other magnetoelectric materials.  
We underscore that our mechanism is quite general and widely expectable, only requiring the Rashba spin-orbit coupling and exchange coupling. 
Our result could potentially offer a promising method of Joule heating--free electric manipulation of magnetic moments in spintronic devices. 
\end{abstract}

\maketitle

\section{Introduction}\label{sec:intro}
Electric generation and control of spin degrees of freedom has been a central issue in the field of spintronics.
For over the decades, a number of researches have been dedicated to manipulation of ferromagnetic moments or domain walls by an electric current in semiconductors and metals towards realization of magnetic memory without need of a magnetic field~\cite{Brataas2012}.
During the time, several kinds of methods have been developed that allow for magnetization control by an electric current.
One of the early examples would be spin transfer torque exerted by a spin-polarized current~\cite{Slonczewski1996, Berger1996}, which, for instance, was demonstrated in the experiment of current-induced domain wall motion in a magnetic semiconductor GaMnAs (see Ref.~\cite{Brataas2012} and the references therein).
In recent years, the Edelstein effect~\cite{Edelstein1990, Aronov1989} and the spin Hall effect~\cite{Murakami2003, Sinova2015} have been gathering much attention as prominent spin-charge conversion methods. 
The former produces nonequilibrium spin accumulation through modulation of the Fermi surface by a current flow in a noncentrosymmetric metal, whereas the latter generates a pure spin current perpendicular to the electric current.
Both two phenomena enable efficient manipulation of ferromagnetic moments through exertion of torque termed spin-orbit torque, which has been confirmed in bilayer thin films composed of a heavy metal and a ferromagnet, e.g., Pt/Co~\cite{Miron2010, Pi2010, Miron2011, Liu2012prl, Liu2012science}.
We should also mention that lately current-driven magnetization control has been extended to antiferromagnetic orders, e.g., in CuMnAs~\cite{Zelezny2014, Wadley2016} or noncollinear orders, e.g., in Mn$_3$Sn~\cite{Zelezny2017}, which might be favored over ferromagnetism in robustness against magnetic perturbations.

As mentioned, the Edelstein effect, or sometimes called the inverse spin galvanic effect, refers to nonequilibrium spin accumulation by a current flow in noncentrosymmetric metals~\cite{Edelstein1990, Aronov1989}. 
The phenomenon is usually understood in the semiclassical picture based on the Boltzmann transport theory~\cite{Silsbee2004}.
In metals with spin-split Fermi surfaces by the relativistic spin-orbit coupling, an electric current generates net spin accumulation through a modification in the momentum distribution function around the Fermi energy, i.e., the shift of the Fermi surfaces~\cite{Silsbee2004, Ando2017}.
The Edelstein effect has been observed in a number of materials, ranging from noncentrosymmetric semiconductors~\cite{Kato2004, Sih2005, Silov2004} and metal heterostructures~\cite{Zhang2015} to topological insulator surfaces~\cite{Mellnik2014}.

Meanwhile, from the quantum-mechanical perspective, the semiclassical picture in the Edelstein effect falls short of the thorough description of the electric field--induced magnetization, because it takes into account only the intraband contributions and omit the interband contributions coming from virtual interband transitions. 
Specifically, when time-reversal symmetry (TRS) is broken in the system, e.g., in proximity to a ferromagnet, the linear response theory indicates that the electric field--induced magnetization includes the intrinsic contributions stemming from the interband virtual transition processes~\cite{Freimuth2014, Li2015, Zelezny2014}.
These intrinsic contributions are immune to disorder or impurity scattering as in the intrinsic anomalous Hall effect~\cite{Nagaosa2010}.
For instance, it has been theoretically pointed out that in two-dimensional Rashba metals coupled with a ferromagnet, these intrinsic components substantially contributes to the total magnetization in a qualitatively distinct manner~\cite{Li2015, Pesin2012}: in the case that the ferromagnetic moment points along the out-of-plane direction (${\mathbf z}$), the magnetization induced by electric field ${\mathbf E}$ is oriented along ${\mathbf z}\times{\mathbf E}$ for the intraband contributions, whereas it is along ${\mathbf E}$ for the intrinsic interband ones. 

These quantum-mechanical descriptions naturally offer a chance that electric field--induced magnetization could be available even in band insulators by relying on the intrinsic contributions.
This current-free magnetoelectric effect in band insulators is distinguished from the kinetic magnetoelectric effects involving flow of an electric current, e.g., the Edelstein effect.
The manifestation of the magnetoelectric effect is restricted by symmetry.
Symmetry arguments dictate that the linear magnetoelectric effect, which is the magnetoelectric response linear to an external electric field, requires breaking of both space-inversion symmetry and TRS~\cite{Fiebig2005}, whereas the Edelstein effect only requires breaking of the former symmetry. 
To date, the magnetoelectric effect in band insulators has been predicted in topological materials showing the quantum anomalous Hall effect~\cite{Hanke2017, Garate2010}.
However, it still remains elusive whether the magnetoelectric effect could commonly appear in band insulators without topological nature. 
From the standpoint of practical applications, the magnetoelectric effect in band insulators would be highly desirable due to the dissipationless nature. This is in stark contrast to the Edelstein effect, where a large amount of Joule heating by an electric current could hamper operations in actual devices.

In this work, we theoretically investigate the magnetoelectric effect in a heterostructure composed of a generic band insulator and a ferromagnet.
Our model of two-dimensional band insulators includes the Rashba spin-orbit coupling stemming from breaking of the mirror symmetry due to the heterostructure as well as the exchange coupling between electrons in the band insulator and the ferromagnetic moment, which fulfills the symmetrical requirements for the linear magnetoelectric effect. 
By calculating the magnetoelectric tensor via the linear response theory, we find that the magnetoelectric effect is generally manifested in a broad range of the parameters in accordance with the inherent symmetry. 
We also reveal that the direction of the electric field--induced magnetization depends on the orientation of the magnetic moment in the ferromagnet, which is also explained by the symmety arguments.
Furthermore, we show that the magnetoelectric effect can be resonantly amplified in optical regions.
Finally, we estimate the magnitude of the magnetoelectric effect for a noncentrosymmetric hybrid halide perovskite (HHP) semiconductor CH$_3$NH$_3$PbI$_3$ in a heterostructure with a ferromagnet and compare it with some known magnetoelectric materials. 

The rest of the paper is orgnaized as follows.
In Sec.~\ref{sec:model}, we introduce the models describing the heterostructures with a ferromagnet for a generic band insulator and a HHP semiconductor. We also represent the Kubo formula for the magnetoelectric coefficient.
In Sec.~\ref{sec:general}, we study the magnetoelectric effect in the generic band insulator.
In Sec.~\ref{sec:magnormal}, we study the case where the magnetic moment points along the out-of-plane direction, whereas in Sec.~\ref{sec:mdd}, we study the case with the moment canted away.
We discuss these results in terms of symmetry in Sec.~\ref{sec:sym}.
Then we investigate the optical magnetoelectric effect for the out-of-plane magnetic moment in Sec.~\ref{sec:ome}.
In Sec.~\ref{sec:hhp}, we focus on a specific example of the magnetoelectric effect discussed in Sec.~\ref{sec:general}, bringing up a HHP semicoductor.
We estimate the magnitude of the magnetoelectric effect expected in the HHP semicoductor in Sec.~\ref{sec:est} and compare it with other known magnetoelectric effects in Sec.~\ref{sec:dis}. 
Finally, Sec.~\ref{sec:summary} is devoted to a summary of our results.

\section{Model}\label{sec:model}
\subsection{Heterostructure of a two-dimensional band insulator and a ferromagnet}\label{sec:toymodel}
\begin{figure}[!htb]
\begin{center}
\includegraphics[width=0.7\columnwidth,clip]{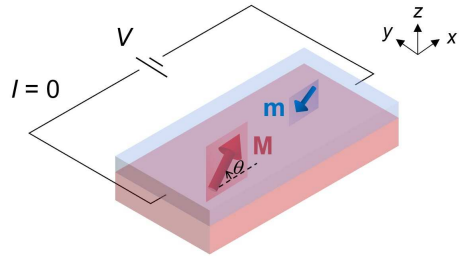}
\end{center}
\caption{Schematic picture of an experimental setup for the magnetoelectric effect on a heterostructure of a two-dimensional band insulator and a ferromagnet. ${\mathbf M}$ represents the magnetic moment of the ferromagnet and ${\mathbf m}$ the magnetization induced by the voltage $V$. $\theta$ represents the angle of the moment ${\mathbf M}$ measured from the $x$ axis in the $zx$ plane. We note that no electric current flows ($I=0$).}
\label{fig:schematic}
\end{figure}

To capture the essential features of the magnetoelectric effect in band insulators, we construct a tight-binding model on a square lattice describing a heterostructure of a two-dimensional band insulator and a ferromagnet, where the Rashba spin-orbit coupling arises due to breaking of the mirror symmetry with respect to the interface.
Here we set the $z$ axis to the out-of-plane direction, while the $x$ and $y$ axes within the plane (see Fig.~\ref{fig:schematic}).
As a minimal model for band insulators, we consider two orbitals, whose wave fuctions are spherical in a similar manner to $s$ orbitals.
The Rashba spin-orbit coupling is introduced through the electron hopping under the inversion-asymmetric potential due to the heterostructure.
The Hamiltonian is given in momentum space as
\begin{equation} 
H({\mathbf k})=\left(
    \begin{array}{cc}
      H_1({\mathbf k})&H_{12}({\mathbf k})\\
      H^\dagger_{12}({\mathbf k})&H_{2}({\mathbf k})\\
    \end{array}
  \right)
\label{eq:toymodel}
\end{equation}
in the basis of $\left\{\ket{1\uparrow}, \ket{1\downarrow}, \ket{2\uparrow}, \ket{2\downarrow}\right\}$, where $1$ and $2$ denote the two orbitals, and $\uparrow$ and $\downarrow$ denote the spins.
The intra and interorbital matrix components $H_\rho(\mathbf{k})$ ($\rho=1$ and $2$) and $H_{12}({\mathbf k})$ are $2\times 2$ matrices spanned in spin space, which are given by 
\begin{align} 
\label{eq:Hintra} 
\begin{split} 
H_\rho({\mathbf k})&=2t_\rho(\cos k_x+\cos k_y)I\\
&\quad+2\lambda_\rho(\sin k_y\sigma_x-\sin k_x\sigma_y)+\eta(\rho)\frac{\epsilon_{12}}{2}I\\
&\quad+J_{\rm ex}(M_x\sigma_x+M_z\sigma_z),
\end{split}\\ 
\label{eq:Hinter} 
\begin{split} 
H_{12}({\mathbf k})&=2t_{12}(\cos k_x+\cos k_y)I\\
&\quad+2\lambda_{12}(\sin k_y\sigma_x-\sin k_x\sigma_y),
\end{split} 
\end{align}
respectively.
We set the lattice constant $a$ to be unity ($a=1$). 
Here $\sigma_i$ ($i=x, y,$ and $z$) is the Pauli matrix, and $I$ is the identity matrix.
$t_\rho$ and $t_{12}$ are the intra and interorbital transfer integrals, respectively; 
$\lambda_\rho$ and $\lambda_{12}$ are the effective intra and interorbital transfer integrals arising from the Rashba spin-orbit coupling, respectively. 
$\epsilon_{12}$ is the energy splitting between the orbitals $1$ and $2$ ($\epsilon_{12}>0$);  
$\eta(\rho)$ is defined as $\eta(1)=-1$ and $\eta(2)=1$.
We define the magnetic moment of the ferromagnet as ${\mathbf M}$, whose length is normalized as $|{\mathbf M}|=1$.
We assume ${\mathbf M}$ within the $zx$ plane, parametrized as ${\mathbf M}=(M_x, 0, M_z)$. 
$J_{\rm ex}$ denotes the exchange coupling between ${\mathbf M}$ and the spins of the electrons in the band insulator. 
The detailed derivation of the model in Eq.~(\ref{eq:toymodel}) is given in Appendix~A. 
In the following calculations, we set $t_1=1$, $t_2=-1$, and $\lambda_1=0.3$.

\subsection{Hybrid halide perovskite semiconductor}\label{sec:HHPmodel}
To estimate the magnitude of the magnetoelectric effect in a real material, we also perform the calculations on a noncentrosymmetric HHP semiconductor CH$_3$NH$_3$PbI$_3$ as the band insulator in the heterostructure.
This material is known to host a large Rashba spin-orbit coupling even in the bulk form, due to the ferroelectricity combined with the strong atomic spin-orbit coupling~\cite{Kim2014}. 
The electronic bands near the chemical potential are mostly composed of $s$ and $p$ orbitals of Pb. 
The atomic spin-orbit coupling splits the $p$ orbitals to $J=\frac{1}{2}$ and $J=\frac{3}{2}$ states, and consequently, the valence and conduction bands are predominantly constructed by the $s$-orbital $S=\frac{1}{2}$ states and the $p$-orbital $J=\frac{1}{2}$ states, respectively.
Using the minimal Hamiltonian for these valence and conduction bands derived in Ref.~\cite{Li2017}, we consider the Hamiltonian for a heterostructure with a ferromagnet as 
\begin{equation} 
H({\mathbf k})=\left(
    \begin{array}{cc}
      H_S({\mathbf k})&H_{SJ}({\mathbf k})\\
      H^\dagger_{SJ}({\mathbf k})&H_{J}({\mathbf k})\\
    \end{array}
  \right)
\label{eq:hhpmodel}
\end{equation}
in the basis of $\left\{\ket{S_z=\frac{1}{2}}, \ket{S_z=-\frac{1}{2}}, \ket{J_z=\frac{1}{2}}, \ket{J_z=-\frac{1}{2}}\right\}$, where $\ket{S_z=\pm\frac{1}{2}}=\ket{s,s_z=\pm\frac{1}{2}}$ and $\ket{J_z=\pm\frac{1}{2}}=-\frac{1}{\sqrt{3}}(\ket{p_x,s_z=\mp\frac{1}{2}}\pm{\rm i}\ket{p_y, s_z=\mp\frac{1}{2}}\pm\ket{p_z, s_z=\pm\frac{1}{2}})$ hold. 
As we are interested in two-dimensional systems, our model assumes a $(001)$-oriented monolayer of CH$_3$NH$_3$PbI$_3$. 
The matrix components in Eq.~(\ref{eq:hhpmodel}) are written down as
\begin{align} 
\label{eq:hhps}
H_S({\mathbf k})&=2t^\sigma_{ss}\left[2-\cos(\tilde{k}_xa)-\cos(\tilde{k}_ya)\right]I-\frac{\Delta}{2}\sigma_z,\\
\label{eq:hhpj}
H_J({\mathbf k})&=2\frac{2t^\pi_{pp}+t^\sigma_{pp}}{3}\left[2-\cos(\tilde{k}_xa)-\cos (\tilde{k}_ya)\right]I\nonumber \\
&\quad +\frac{4}{3}\gamma^z_{pp}\left[\sin(\tilde{k}_ya)\sigma_x-\sin(\tilde{k}_xa)\sigma_y\right]+\epsilon_0I+\frac{\Delta}{6}\sigma_z,\\
H_{SJ}({\mathbf k})&=\frac{2}{\sqrt{3}}\gamma^z_{sp}\left[\cos(\tilde{k}_xa)+\cos(\tilde{k}_ya)\right]\sigma_z\nonumber \\
&\quad +{\rm i}\frac{2}{\sqrt{3}}t^\sigma_{sp}\left[\sin(\tilde{k}_xa)\sigma_x+\sin(\tilde{k}_ya)\sigma_y\right],
\end{align}
where $\tilde{{\mathbf k}}=(\tilde{k}_x, \tilde{k}_y)$ is the momentum measured from the M point, represented as $\tilde{k}_i=k_i-\pi/a$. 
Here $t^\sigma_{ss}$, $t^\sigma_{pp}$, and $t^\pi_{pp}$ represent the transfer integrals for the $\sigma$-type hopping between the $s$ orbitals and $\sigma$- and $\pi$-type hoppings between the $p$ orbitals; $t^\sigma_{sp}$ is the interorbital transfer integral, and $\gamma^z_{sp}$ and $\gamma^z_{pp}$ are the transfer integrals activated by the mirror-symmetry breaking. 
$\epsilon_0$ represents the energy of the $J=\frac{1}{2}$ states measured from that of the $S=\frac{1}{2}$ states.
Following Ref.~\cite{Li2017}, we set $t^\sigma_{ss}=-0.25$~eV, $t^\sigma_{pp}=0.9$~eV, $t^\pi_{pp}=0.15$~eV, $t^\sigma_{sp}=0.4$~eV, $\gamma^z_{pp}=-0.2$~eV, $\gamma^z_{sp}=-0.25$~eV, and $\epsilon_0=1.5$~eV.
In Eqs.~(\ref{eq:hhps}) and (\ref{eq:hhpj}), $\Delta$ represents an exchange splitting arising from the coupling to the out-of-plane magnetic moment in the adjacent ferromagnet.
Note that the exchange splitting in the $J=\frac{1}{2}$ band has the opposite sign and one third of the magnitude relative to that in the $S=\frac{1}{2}$ bands, due to the spin-orbit--coupled nature of the $J=\frac{1}{2}$ pseudospin [see also Eq.~(\ref{eq:HHPSigma})]. 

\subsection{Magnetoelectric tensor}\label{sec:me}
To calculate the response coefficient of the magnetoelectric effect, we employ the Kubo formula for the magnetoelectric tensor.
In the linear response regime, the magnetoelectric tensor $\alpha_{ij}$ is defined as
\begin{equation} 
m_i=\alpha_{ij}E_j,
\label{eq:me}
\end{equation}
where $m_i$ and $E_j$ denote the $i$ component of the induced magnetization and $j$ component of the external electric field.
The magnetoelectric tensor $\alpha_{ij}$ is obtained by using the Kubo formula as~\cite{Li2015,Hayami2014}
\begin{equation}
\alpha_{ij}=-\mu_{\rm B}K_{ij},
\label{eq:alpha}
\end{equation}
where $\mu_{\rm B}$ is the Bohr magneton ($g$-factor is set as $g=2$), and $K_{ij}$ is given by 
\begin{align}
K_{ij}=-\frac{e}{V}\sum_{\mathbf{k}}\sum_{m\neq n}
&\frac{f(\epsilon_{n\mathbf{k}})-f(\epsilon_{m\mathbf{k}})}{\epsilon_{n\mathbf{k}}-\epsilon_{m\mathbf{k}}}\nonumber \\
\times&\frac{{\rm Im}[\braket{u_{n\mathbf{k}}|\Sigma_i|u_{m\mathbf{k}}}\braket{u_{m\mathbf{k}}|v_j|u_{n\mathbf{k}}}]}{\epsilon_{n\mathbf{k}}-\epsilon_{m\mathbf{k}}}. 
\label{eq:K} 
\end{align}
Here $\epsilon_{n\mathbf{k}}$ and $\ket{u_{n\mathbf{k}}}$ are the eigenenergy and eigenstate of $H(\mathbf{k})$ with the band index $n$.
$e$ denotes the elementary charge, and $V=(La)^2$ is the volume ($L$: the linear dimension of the system, $a$: the lattice constant). 
$f(\epsilon_{n\mathbf{k}})$ is the Fermi distribution function: $f(\epsilon_{n\mathbf{k}})=1/\left[1+{\rm exp}\left\{(\epsilon_{n\mathbf{k}}-\mu)/k_{\rm B}T\right\}\right]$, where $\mu$ and $T$ represent the chemical potential and temperature, respectively ($k_{\rm B}$: the Boltzmann constant).
The velocity operator $v_i$ is defined as $v_i=\partial H(\mathbf{k})/\partial k_i$.
We note that since we treat band insulators, intraband contributions vanish in the summation of the band indices in Eq.~(\ref{eq:K}). 
In the calculations on the generic model in Eq.~(\ref{eq:toymodel}), we set $e=k_{\rm B}=1$.
We set almost zero temperature, where the smearing effect of the Fermi distribution function is negligible. 

The spin operator $\Sigma_i$ in Eq.~(\ref{eq:K}) depends on the model. 
In the generic model introduced in Sec.~\ref{sec:toymodel}, $\Sigma_i$ is simply composed of the Pauli matrices $\sigma_i$ as
\begin{equation} 
\Sigma_i=\left(
    \begin{array}{cc}
      \sigma_i&0\\
      0&\sigma_i\\
    \end{array}
  \right)
\end{equation}
in the same basis as in Eq.~(\ref{eq:toymodel}). 
Meanwhile, in the model for the HHP semiconductor in Sec.~\ref{sec:HHPmodel}, $\Sigma_i$ is given in the form of 
\begin{equation} 
\Sigma_i=\left(
    \begin{array}{cc}
      \sigma_i&0\\
      0&-\sigma_i/3\\
    \end{array}
  \right)
\label{eq:HHPSigma} 
\end{equation} 
in the same basis as in Eq.~(\ref{eq:hhpmodel}), owing to the nature of the pseudospin in the $J=\frac12$ states.

We also study the optical magnetoelectric response on the model in Eq.~(\ref{eq:toymodel}) in Sec.~\ref{sec:ome}. We calculate the finite-frequency version of $K_{ij}$ in Eq.~(\ref{eq:K}), which is given by 
\begin{align}
K_{ij}(\omega)=-\frac{e}{{\rm i}V}\sum_{\mathbf{k}}\sum_{m\neq n}
&\frac{f(\epsilon_{n\mathbf{k}})-f(\epsilon_{m\mathbf{k}})}{\epsilon_{n\mathbf{k}}-\epsilon_{m\mathbf{k}}}\nonumber \\
\times&\frac{\braket{u_{n\mathbf{k}}|\Sigma_i|u_{m\mathbf{k}}}\braket{u_{m\mathbf{k}}|v_j|u_{n\mathbf{k}}}}{\epsilon_{n\mathbf{k}}-\epsilon_{m\mathbf{k}}+\hbar\omega+{\rm i}\delta}. 
\label{eq:Komega} 
\end{align}

\section{Magnetoelectric effect in a general band insulator}\label{sec:general}
\subsection{Case with out-of-plane magnetic moment}\label{sec:magnormal}
\begin{figure}[!htb]
\centering
\includegraphics[width=\columnwidth,clip]{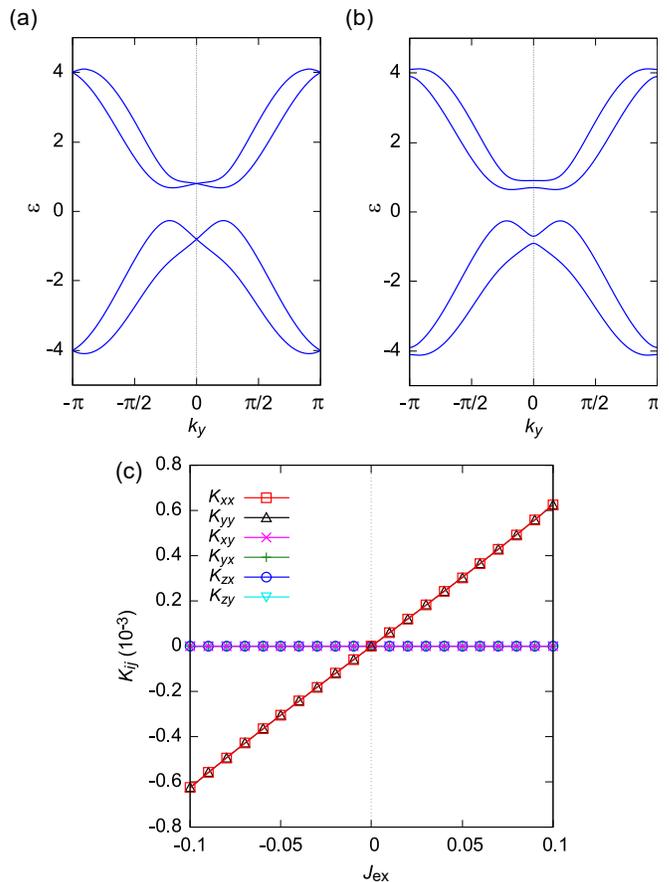}
\caption{Band structures for $k_x=0$ (a) without ($J_{\rm ex}=0$) and (b) with the exchange coupling to the out-of-plane magnetic moment ${\mathbf M}=(0, 0, 1)$ ($J_{\rm ex}=0.1$). 
(c) $K_{ij}$ as a function of $J_{\rm ex}$ for ${\mathbf M}=(0, 0, 1)$. We set the model parameters as $t_1=1$, $\lambda_1=0.3$, $t_2=-1$, $\lambda_2=0.3$, $\epsilon_{12}=8$, $t_{12}=-0.2$, and $\lambda_{12}=0.2$.}
\label{fig:K_Jz}
\end{figure}

In this section, we discuss the magnetoelectric effect when the magnetic moment ${\mathbf M}$ points along the $z$ axis [${\mathbf M}=(0, 0, 1)$]. 
We set the model parameters as $\lambda_2=0.3$, $t_{12}=-0.2$, $\lambda_{12}=0.2$, and $\epsilon_{12}=8$ in Eqs.~(\ref{eq:Hintra}) and (\ref{eq:Hinter}), while the following results are qualitatively the same in a wide region of the parameters.
In Figs.~\ref{fig:K_Jz}(a) and \ref{fig:K_Jz}(b), we present the band structures for $k_x=0$ without ($J_{\rm ex}=0$) and with the exchange coupling to ${\mathbf M}$ ($J_{\rm ex}=0.1$), respectively.
Figure~\ref{fig:K_Jz}(a) shows that in the absence of the exchange coupling, the valence and conduction bands are separated by a substantial band gap ($E_{\rm g}\sim 0.9$), and each band exhibits spin splitting by the Rashba spin-orbit couplings $\alpha_\rho$ and $\alpha_{12}$.
The spin degeneracy is lifted over the whole Brillouin zone (BZ), except the time-reversal invariant momenta (TRIMs), namely, $\Gamma=(0, 0)$, ${\rm X}_1=(\pi, 0)$, ${\rm X}_2=(0, \pi)$, and ${\rm M}=(\pi, \pi)$. 
At each TRIM, a pair of bands show a linear crossing both in the valence and conduction bands, creating two Dirac points as shown in Fig.~\ref{fig:K_Jz}(a). 
When the out-of-plane magnetic moment ${\mathbf M}$ is introduced, gap opening occurs at the Dirac points as shown in Fig.~\ref{fig:K_Jz}(b).

\begin{figure}[!htb]
\begin{center}
\includegraphics[width=0.7\columnwidth,clip]{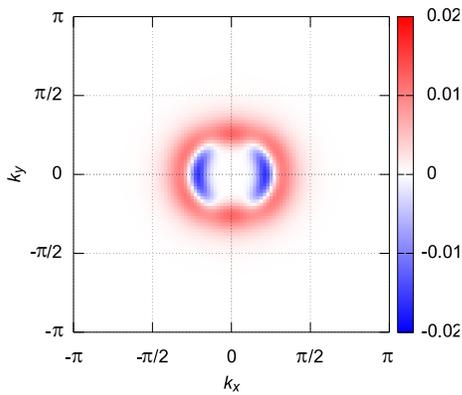}
\end{center}
\caption{Momentum-resolved $K_{xx}$ with $J_{\rm ex}=0.1$ and ${\mathbf M}=(0, 0, 1)$. We use the same model parameters as in Fig.~\ref{fig:K_Jz}.}
\label{fig:K_kprofile}
\end{figure}

In Fig.~\ref{fig:K_Jz}(c), we present $J_{\rm ex}$ dependence of the coefficient $K_{ij}$ obtained by Eq.~(\ref{eq:K}). 
In the calculations, we set the chemical potential $\mu$ within the band gap ($\mu\sim0$). 
Figure~\ref{fig:K_Jz}(c) shows that without the exchange coupling ($J_{\rm ex}=0$), all the elements of $K_{ij}$ vanish, which is imposed by TRS.
As $J_{\rm ex}$ increases from zero, the diagonal components $K_{xx}$ and $K_{yy}$ develop linearly with $J_{\rm ex}$, which marks the diagonal magnetoelectric effect. 
On the other hand, the off-diagonal components always remain zero, which will be accounted for by the symmetry arguments in Sec.~\ref{sec:sym}.
We also see that as $J_{\rm ex}$ turns into a negative value, which is equivalent to ${\mathbf M}$ reversal with fixed $J_{\rm ex}$, the sign of $K_{xx}$ and $K_{yy}$ changes with the identical magnitude.
The sign change in $K_{ij}$ by the time-reversal operation, i.e., the flip of the moment ${\mathbf M}$, can be derived from the transformation of the Hamiltonian by the time-reversal operator $\Theta={\rm i}\Sigma_yK$ ($K$: complex conjugate operator): $\Theta H({\mathbf k};{\mathbf M})\Theta^{-1}=H(-{\mathbf k};-{\mathbf M})$.
We remark that the induced magnetization ${\mathbf m}$ can exert a torque ${\mathbf T}\propto{\mathbf m}\times{\mathbf M}$ against the moment ${\mathbf M}$~\cite{Li2015}: for ${\mathbf E}||{\mathbf x}$, ${\mathbf m}\propto{\mathbf x}$ and ${\mathbf T}\propto{\mathbf y}$. 

Next we examine the observed magnetoelectric effect microscopically.
In Fig.~\ref{fig:K_kprofile}, we display momentum-resolved $K_{xx}$ for $J_{\rm ex}=0.1$, whose sum over the BZ amounts to $K_{xx}$ in Fig.~\ref{fig:K_Jz}(c).
The result indicates that only the region around the ${\rm\Gamma}$ point is responsible for $K_{xx}$.  
This stems from the fact that orbital mixing and accordingly the spin matrix element $\braket{u_{n\mathbf{k}}|\Sigma_x|u_{m\mathbf{k}}}$ is substantial only in the vicinity of the ${\rm\Gamma}$ point.
Therefore, the magnetoelectric effect stems from the virtual interband transitions between the two massive Rashba dispersions in the valence and conduction bands around the ${\rm\Gamma}$ point.

Meanwhile, the observed magnetoelectric effect can also be viewed from a topological perspective.
In analogy to the Thouless--Kohmoto--Nightingale--den Nijs (TKNN) formula for the Hall conductance~\cite{Thouless1982}, the linear response formula for the magnetoelectric coefficient $\alpha_{ij}$ in Eqs.~(\ref{eq:alpha}) and (\ref{eq:K}) is also written as~\cite{Hanke2017} 
\begin{align}
\alpha_{ij}=\frac{e\mu_{\rm B}}{J_{\rm ex}}\frac{1}{2\pi^2}\sum_{n:\,{\rm occ}}\int dk_xdk_y\tilde{\Omega}^{ij}_{n\mathbf{k}}.
\label{eq:alpha_berry}
\end{align}
Here $\tilde{\Omega}^{ij}_{n\mathbf{k}}$, termed the mixed Berry curvature~\cite{Hanke2017}, represents the Berry curvature defined on the magnetic moment--momentum--hybrid $M_ik_j$ plane, which is formulated as $\tilde{\Omega}^{ij}_{n\mathbf{k}}=2{\rm Im}[\braket{\partial_{M_i}u_{n{\mathbf k}}|\partial_{k_j}u_{n{\mathbf k}}}]$.
Hence, the formalism in Eq.~(\ref{eq:alpha_berry}) indicates that the momentum-dependent magnetoelectric signal shown in Fig.~\ref{fig:K_kprofile} can also be regarded as the momentum profile of the mixed Berry curvature. 
Later in Sec.~\ref{sec:dis}, we discuss the implications from this topological viewpoint.

\begin{figure}[!htb]
\begin{center}
\includegraphics[width=\columnwidth,clip]{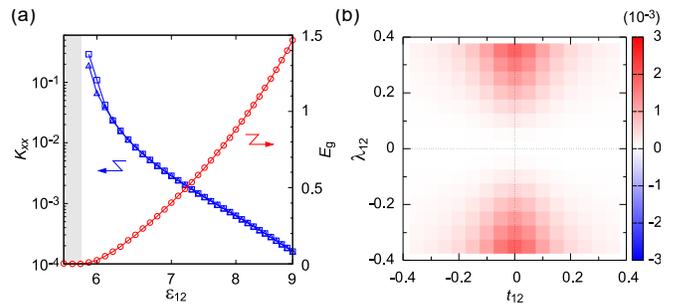}
\end{center}
\caption{(a) $\epsilon_{12}$ dependence of $K_{xx}$ and $E_{\rm g}$. We plot $K_{xx}$ for $L=96$ (squares) and $L=480$ (triangles). The grey-shaded area corresponds to a metallic region. (b) $t_{12}$-$\lambda_{12}$ dependence of $K_{xx}$. We set $J_{\rm ex}=0.1$ and ${\mathbf M}=(0, 0, 1)$ and use the same model parameters as in Fig.~\ref{fig:K_Jz} except $\epsilon_{12}=9$ in (b).}
\label{fig:K_e12}
\end{figure}

Hitherto, we have focused on a particular set of the model parameters, but as we describe in the following, the magnetoelectric effect generally manifests itself in a broad range of the parameters.
In Fig.~\ref{fig:K_e12}(a), we present $\epsilon_{12}$ dependence of $K_{xx}$ for $J_{\rm ex}=0.1$ together with the band gap $E_{\rm g}$. 
We use the same model parameters as in Fig.~\ref{fig:K_Jz}.
As $\epsilon_{12}$ decreases, $E_{\rm g}$ shrinks and eventually collapses at $
\epsilon_{12}\sim 5.8$, where the system turns into a metal [see the grey-shaded area in Fig.~\ref{fig:K_e12}(a)].
In the meantime, $K_{xx}$ grows with the decrease in $\epsilon_{12}$ and shows nearly-diverging behavior near the insulator-metal transition. 
The enhancement of $K_{xx}$ can be readily attributed to the decrease in the energy difference in the denominator in Eq.~(\ref{eq:K}).

We also study the dependence of the magnetoelectric effect on the interband parameters $t_{12}$ and $\lambda_{12}$.
Figure~\ref{fig:K_e12}(b) displays the results of $K_{xx}$ for $J_{\rm ex}=0.1$.
The other model parameters are the same as in Fig.~\ref{fig:K_Jz} except $\epsilon_{12}$; we set $\epsilon_{12}=9$ just to keep the band gap open in the entire range of $t_{12}$ and $\lambda_{12}$ in Fig.~\ref{fig:K_e12}(b).
When $t_{12}=\lambda_{12}=0$, $K_{xx}$ is obviously zero, because the two orbitals are not mixed at all.
Figure~\ref{fig:K_e12}(b) indicates that the interorbital Rashba spin-orbit coupling $\lambda_{12}$ is important to $K_{xx}$, where the increase in $|\lambda_{12}|$ magnifies $K_{xx}$.
We note that the decrease in $|t_{12}|$ leads to enhancement of $K_{xx}$, which can be attributed to shrinkage of the band gap.
In Appendix B, we also discuss $\lambda_2$ dependence and find out that the magnetoelectric effect appears for general values of $\lambda_2$, other than the case with $\lambda_2=-\lambda_1$ characterized with particle-hole symmetry.
Thus, the magnetoelectric effect generally appears without specific tuning of the parameters in the presence of the Rashba spin-orbit coupling and the exchange coupling, and the signal is roughly proportional to both coupling strengths. 

\subsection{Dependence on the magnetic moment direction}\label{sec:mdd}
\begin{figure}[!htb]
\begin{center}
\includegraphics[width=\columnwidth,clip]{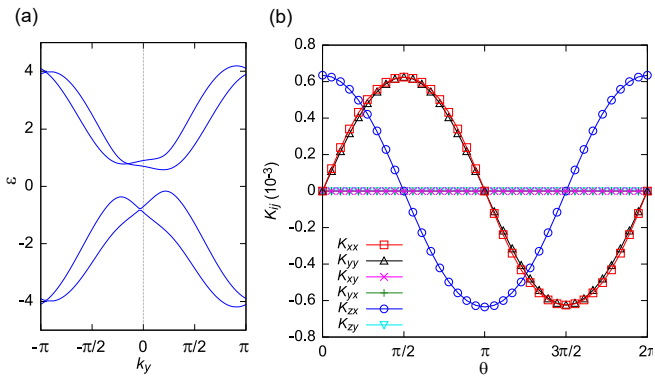}
\end{center}
\caption{(a) Band structure for $k_x=0$ with $\theta=0$. $\theta$ is the angle of the ${\mathbf M}$ direction, defined as $M_x=\cos\theta$ and $M_z=\sin\theta$ (see also Fig.~\ref{fig:schematic}). (b) $\theta$ dependence of $K_{ij}$. We set $J_{\rm ex}=0.1$ and use the same model parameters as in Fig.~\ref{fig:K_Jz}.}
\label{fig:K_q}
\end{figure}

In this section, we discuss how the magnetoelectric effect in the band insulator evolves as the ${\mathbf M}$ direction changes in the ferromagnet.
We consider a situation where ${\mathbf M}$ rotates in the $zx$ plane with $J_{\rm ex}=0.1$.
Here we parametrize the ${\mathbf M}$ direction with the angle $\theta$, which is defined as ${\mathbf M}=(M_x, 0, M_z)=(\cos\theta, 0, \sin\theta)$.
The other model parameters are identical to those in Fig.~\ref{fig:K_Jz}.
As already stated in Sec.~\ref{sec:magnormal}, when ${\mathbf M}$ points along the $z$ axis ($\theta=\pi/2$), the Dirac points acquire mass gaps at the four TRIMs [see Fig.~\ref{fig:K_q}(a)]. 
On the other hand, as ${\mathbf M}$ is canted to the $x$ axis, the Dirac points are shifted along the $k_y$ direction without being gapped out.
As shown in Fig.~\ref{fig:K_q}(a), the Dirac points are shifted in the same direction between the valence and conduction bands, due to the same sign of the Rashba spin-orbit couplings $\lambda_1$ and $\lambda_2$.  

In Fig.~\ref{fig:K_q}(b), we present the $\theta$ dependence of $K_{ij}$.
The result reveals that $K_{xx}$ and $K_{yy}$ evolve nearly in proportion to $M_z=\sin\theta$, whereas $K_{zx}$ in proportion to $M_x=\cos\theta$.
In particular, when ${\mathbf M}$ points along the $z$ axis, ${\mathbf E}||{\mathbf x}$ induces ${\mathbf m}||{\mathbf x}$ as already studied in Sec.~\ref{sec:magnormal}, whereas when ${\mathbf M}$ is canted to the $x$ axis, ${\mathbf E}||{\mathbf x}$ induces ${\mathbf m}||{\mathbf z}$. 
On the other hand, $K_{xy}$, $K_{yx}$, and $K_{zy}$ remain inactivated regardless of the angle $\theta$. 
We note that as explained in Sec.~\ref{sec:magnormal}, $K_{ij}$ changes its sign associated with ${\mathbf M}$ reversal, i.e., $\theta\rightarrow\theta+\pi$, as required by the time-reversal operation. 

\subsection{Symmetry argument}\label{sec:sym}
In this section, we qualitatively discuss the dependence on the $\mathbf{M}$ direction in terms of symmetry.
In general, the form of the magnetoelectric tensor is restricted by symmetry inherent in the system. 
As often discussed in the context of multiferroics, the tensor form is decided by symmetry considerations on the electromagnetic free energy that is expanded with respect to the electric field $\mathbf{E}$ and the magnetic field ${\mathbf H}$~\cite{Fiebig2005}.
In the expansion of the free energy, the term describing linear magnetoelectric coupling is represented as 
\begin{align}
f=-\alpha_{ij}H_iE_j,
\label{eq:f}
\end{align}
where $\alpha_{ij}$ denotes the magnetoelectric tensor in Eq.~(\ref{eq:me}). 

In the following, we specify the form of $\alpha_{ij}$ based on the symmetry inherent in the heterostructure. 
We denote the time-reversal operation as $\mathcal{T}$, the mirror operation about the plane normal to the $\xi$ axis as $\mathcal{M}_\xi$ ($\xi=x$ and $y$), and the $2\pi /p$ rotation about the $z$ axis as $\mathcal{C}_p$ ($p$: integer).
As a common feature to all $\theta$, the system retains the combined symmetry $\mathcal{M}_y\mathcal{T}$, because ${\mathbf M}$ lies in the $zx$ plane.
Under the symmetry operation $\mathcal{M}_y\mathcal{T}$, the electromagnetic fields are transformed as
\begin{align}
\begin{split}
\mathcal{M}_y\mathcal{T}:&\ E_x\rightarrow E_x,\ E_y\rightarrow -E_y,\ E_z\rightarrow E_z,\\
&\ H_x\rightarrow H_x,\ H_y\rightarrow -H_y,\ H_z\rightarrow H_z,
\label{eq:transform}
\end{split}
\end{align}
since ${\mathbf E}$ is a polar vector, whereas ${\mathbf H}$ is an axial vector.
As the free energy is unchanged under the symmetry operation, the transformations lead to $\alpha_{xy}=\alpha_{yx}=\alpha_{zy}=\alpha_{yz}=0$. 
This is consistent with the numerical calculations shown in Fig.~\ref{fig:K_q}(b).
We note that $\alpha_{iz}=0$ ($i=x,y,z$) by definition in Eq.~(\ref{eq:K}).

We further dig into two specific cases of the ${\mathbf M}$ direction.
When ${\mathbf M}$ points along the $z$ axis ($\theta=\pi/2$), the heterostructure is characterized with two additional symmetries $\mathcal{M}_x\mathcal{T}$ and ${\mathcal C}_4$. 
By considering the transformations similar to Eq.~(\ref{eq:transform}), we find that these symmetries dictate $\alpha_{zx}=\alpha_{xz}=0$ and $\alpha_{xx}=\alpha_{yy}$, respectively, resulting in an isotropic diagonal magnetoelectric effect as shown in Figs.~\ref{fig:K_Jz}(c) and \ref{fig:K_q}(b).
On the other hand, when ${\mathbf M}$ lies along the $x$ axis ($\theta=0$), the heterostructure is constrained by another extra symmetry $\mathcal{M}_x$.
This, in turn, requires vanishing diagonal elements $\alpha_{xx}=\alpha_{yy}=\alpha_{zz}=0$, which leads to the off-diagonal magnetoelectric effect ($\alpha_{zx}\neq0$) as shown in Fig.~\ref{fig:K_q}(b).
Thus, the qualitative feature of the magnetoelectric effect can also be captured in terms of symmetry. 

\subsection{Optical magnetoelectric responses}\label{sec:ome}
\begin{figure}[!htb]
\begin{center}
\includegraphics[width=0.8\columnwidth,clip]{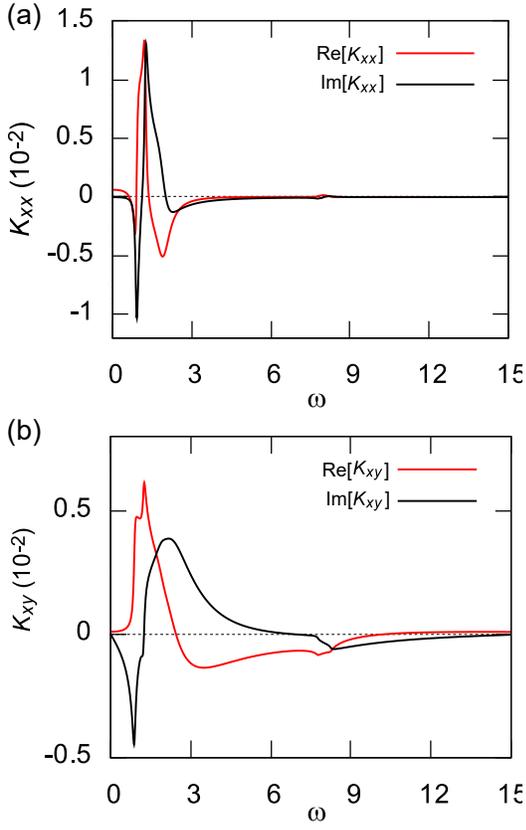}
\end{center}
\caption{Frequency dependence of (a) $K_{xx}$ and (b) $K_{xy}$ with $J_{\rm ex}=0.1$ and ${\mathbf M}=(0, 0, 1)$. 
The red and black curves denote the real and imaginary parts, respectively. 
We use the same model parameters as in Fig.~\ref{fig:K_Jz} and set $\delta=0.05$ in Eq.~(\ref{eq:Komega}).}
\label{fig:Komega}
\end{figure}

In this section, we study the optical magnetoelectric responses in the generic band insulator model in Eq.~(\ref{eq:toymodel}), extending the d.c. result in Fig.~\ref{fig:K_Jz}(c) to finite frequencies.
We set $J_{\rm ex}=0.1$ and ${\mathbf M}=(0, 0, 1)$ and use the same model parameters as in Fig.~\ref{fig:K_Jz}.
We calculate the optical magnetoelectric coefficient $K_{ij}(\omega)$ via Eq.~(\ref{eq:Komega}), setting the smearing factor as $\delta=0.05$.
In Figs.~\ref{fig:Komega}(a) and \ref{fig:Komega}(b), we present the real and imaginary parts for $K_{xx}(\omega)$ and $K_{xy}(\omega)$, respectively.
In contrast to the d.c. result in Fig.~\ref{fig:K_Jz}(c), the off-diagonal component $K_{xy}$ becomes nonzero for finite frequencies as shown in Fig.~\ref{fig:Komega}(b). 
This does not contradict the symmetry arguments in Sec.~\ref{sec:sym}, which no longer hold for the optical regions.
As seen in Fig.~\ref{fig:Komega}(a), $K_{xx}$ is drastically enhanced around the frequency corresponding to the band gap ($E_{\rm g}\sim0.9$), reaching a more than ten times larger value than the d.c. result. 

\section{Magnetoelectric effect in a hybrid halide perovskite semiconductor}\label{sec:hhp}
\subsection{Estimations for out-of-plane magnetic moment}\label{sec:est}
\begin{figure}[!htb]
\begin{center}
\includegraphics[width=\columnwidth,clip]{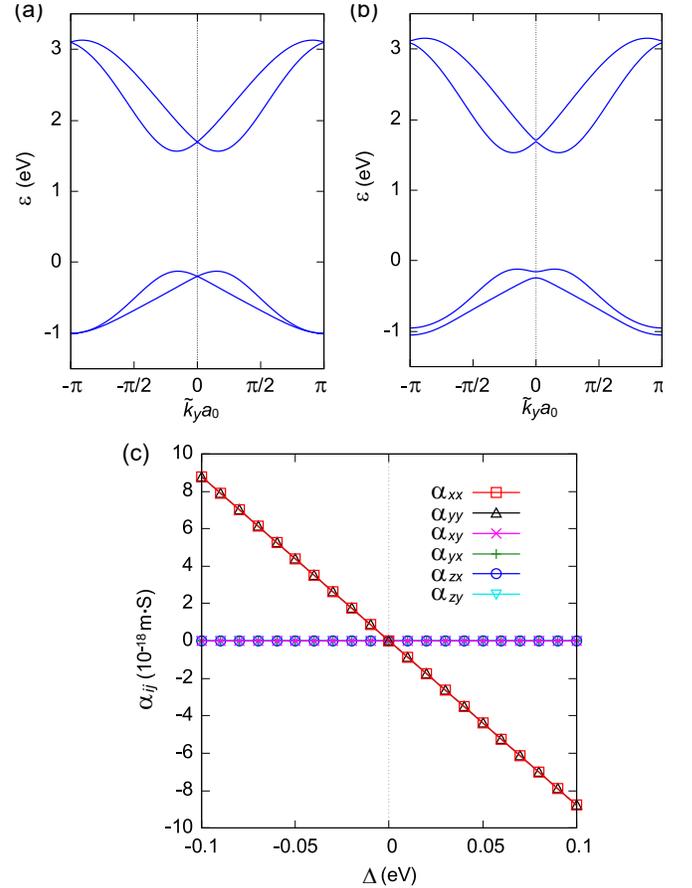}
\end{center}
\caption{Band structures for $\tilde{k}_x=0$ (a) without ($\Delta=0$~eV) and (b) with exchange coupling to the out-of-plane magnetic moment ($\Delta=0.1$~eV) in a HHP semiconductor CH$_3$NH$_3$PbI$_3$. $\tilde{{\mathbf k}}$ is the momentum measured from the M point as $\tilde{k}_i=k_i-\pi/a$. (c) Magnetoelectric coefficient $\alpha_{ij}$ as a function of $\Delta$.}
\label{fig:hhp}
\end{figure}

In the previous sections, we have discussed the magnetoelectric effect on the generic model in Eq.~(\ref{eq:toymodel}).
In this section, we estimate the magnitude of the magnetoelectric effect by taking as an example a noncentrosymmetric HHP semiconductor.
HHP semiconductors CH$_3$NH$_3MX_3$ ($M=$ Pb and Sn; $X=$ I and Br) have recently attracted much attention due to the high solar cell efficiency~\cite{Lee2012}. 
As mentioned in Sec.~\ref{sec:HHPmodel}, these compounds are known to host a large Rashba spin-orbit coupling~\cite{Kim2014}.
As well as the generic model that we have treated in the previous sections, we assume a heterostructure with a ferromagnet for a monolayer form of a HHP semiconductor CH$_3$NH$_3$PbI$_3$. 
In Figs.~\ref{fig:hhp}(a) and \ref{fig:hhp}(b), we present the band structures around the M point without ($\Delta=0$~eV) and with the exchange coupling to the out-of-plane magnetic moment ($\Delta=0.1$~eV).
As shown in Fig.~\ref{fig:hhp}(a), without the exchange coupling, the system has a band gap of $\sim1.7$~eV in the vicinity of the M point, around which the valence and conduction bands show substantial spin splitting by the Rashba spin-orbit coupling.
In the presence of the exchange coupling, the Dirac points are gapped out in a similar way to the previous generic model [see Figs.~\ref{fig:K_Jz}(b) and \ref{fig:hhp}(b)].
We note that the smaller mass gap in the conduction bands is attributed to the $J=\frac{1}{2}$ nature [see Eq.~(\ref{eq:HHPSigma})].

Figure~\ref{fig:hhp}(c) shows the calculated magnetoelectric coefficient $\alpha_{ij}$ in Eq.~(\ref{eq:alpha}) as a function of $\Delta$.
As indicated in Fig.~\ref{fig:hhp}(c), the HHP semiconductor shows the diagonal magnetoelectric effect as well as the previous generic model [see Fig.~\ref{fig:K_Jz}(c)], because both systems are characterized with the same symmetry.
For $\Delta=0.1$ eV, the HHP semiconductor yields the magnetoelectric coefficient of $|\alpha_{xx}|\sim8.7\times 10^{-18}$ m$\cdot$S.  

\subsection{Discussions}\label{sec:dis}
In this section, we compare the results in Sec.~\ref{sec:est} with several other kinds of the magnetoelectric effect. 
As we introduced in Sec.~\ref{sec:intro}, the large magnetoelectric effect has been predicted in quantum anomalous Hall insulators~\cite{Garate2010, Hanke2017}.
Garate $et\ al.$~\cite{Garate2010} predicted that in a magnetic topological insulator, which is the only example where the quantum anomalous Hall effect has been experimentally confirmed~\cite{Chang2013}, the massive Dirac surface states show $\alpha_{xx}=\mu_{\rm B}e/(2hv_{\rm F})$ ($v_{\rm F}$: Fermi velocity of the linear Dirac dispersion) for one side of the surface. This is estimated as $\alpha_{xx}\sim2.8\times 10^{-15}$~m$\cdot$S for a typical value of $v_{\rm F}$ ($v_{\rm F}\sim4.0\times 10^{5}$~m/s~\cite{Zhang2011}).
Hanke $et\ al.$~\cite{Hanke2017} predicted similar magnitudes of the magnetoelectric effect in other quantum anomalous Hall systems by first-principles calculations: for instance, in the presence of exchange coupling with the out-of-plane magnetic moment, a GaBi monolayer would develop $\alpha_{xx}\sim 6.4\times 10^{-15}$~m$\cdot$S associated with the quantum anomalous Hall effect~\footnote{In Ref.~\cite{Hanke2017}, the torkance $\tau_{yx}$, which is the coefficient of the spin-orbit torque induced by electric field, is presented instead of the magnetoelectric coefficient $\alpha_{xx}$. Hence we deduce $\alpha_{xx}$ from $\tau_{yx}$ via the relation $\tau_{yx}=\frac{J_{\rm ex}}{\mu_{\rm B}N_M}\alpha_{xx}$, where $N_M$ represents the number of the localized moments per unit area~\cite{Li2015,Hanke2017}.}.
In comparison with these quantum anomalous Hall systems, the magnetoelectric effect in the HHP semiconductor shown in Fig.~\ref{fig:hhp}(c) remains about three orders of magnitude smaller. 

The large difference can be grasped in terms of the mixed Berry curvature, which was introduced in Sec.~\ref{sec:magnormal}. 
As implicated from the Berry-curvature formalism of the magnetoelectric coefficient in Eq.~(\ref{eq:alpha_berry}), Hanke $et$ $al.$~\cite{Hanke2017} found that $\alpha_{xx}$ can be amplified when the Weyl point exists in the vicinity of the $k_xk_y$ plane in the hybrid three-dimensional $k_xk_yM_x$ space, which is named the mixed Weyl point.
Around the mixed Weyl point, the mixed Berry curvature diverges, just in the same way as the Berry curvature diverges around the Weyl point in the Weyl semimetal~\cite{Yan2017, Armitage2018}.
In the quantum anomalous Hall systems~\cite{Garate2010, Hanke2017}, the mixed Weyl point exists in the vicinity of the $k_xk_y$ plane, because the gap closing occurs by canting the magnetic moment from the $z$ axis to the $x$ axis.
On the other hand, in our models on a generic band insulator as well as the HHP semiconductor, the energy gap is robustly preserved while rotating the magnetic moment, which means the absence of the mixed Weyl point in the hybrid $k_xk_yM_x$ space. 
Therefore, the large separation in the magnitude of the magnetoelectric signal can be qualitatively explained by considering the absence/presence of the mixed Weyl point in the magnetic moment--momentum--hybrid parameter space. 

Although our result is small as compared to the quantum anomalous Hall systems, it should be worth noting that the band gap of the HHP semiconductor ($\sim 1.7$~eV) is more than ten times larger than those in the quantum anomalous Hall systems (tens or hundreds meV)~\cite{Garate2010, Hanke2017}.
This means that in the HHP semiconductor, the insulating nature is much more robust against temperature and unintentional doping by impurities or disorders.
Moreover, as implied by the band gap dependence in Fig.~\ref{fig:K_e12}(a), there is a realistic chance that in some semiconductors with a narrower gap employed in the heterostructure, the magnetoelectric effect could be enhanced by orders of magnitude.

For reference, we also compare our result with the kinetic magnetoelectric effect in metals.
In terms of the magnetoelectric coefficient, the magnitude of the Edelstein effect in typical two-dimensional Rashba metals is the order of $10^{-13}-10^{-12}$~m$\cdot$S~(see Ref.~\cite{Johansson2018} and the references therein).
Thus, the known values of the magnetoelectric effect in band insulators, including our results and the previous works in the quantum anomalous Hall systems~\cite{Garate2010, Hanke2017}, are smaller by orders of magnitude than those of the Edelstein effect.
As stated in Sec.~\ref{sec:intro}, however, the latter involves current flow and should be essentially distinguished from the former. 

Finally, it is also insightful to compare our results with the linear magnetoelectric effect in multiferroic materials.
Cr$_2$O$_3$ is one of the oldest examples of the multiferroic materials~\cite{Fiebig2005}. 
The compound shows a collinear antiferromagnetic order and known to show the linear magnetoelectric effect of $\alpha_{xx}\sim0.3$~$\mu$S via a shift of Cr$^{3+}$ ions.
Through the revival of the magnetoelectric effect, a variety of multiferroic compounds with a much larger linear magnetoelectric effect have been discovered, including TbPO$_4$ ($\alpha\sim240$~$\mu$S)~\cite{Rado1984} and (Fe,Zn)$_2$Mo$_3$O$_8$ ($\alpha\sim88$~$\mu$S)~\cite{Kurumaji2015}. 
By dividing our result by the lattice constant $a\sim0.63$~nm~\cite{Oku2016}, we can effectively derive the magnetoelectric coefficient in the bulk form. 
This results in $\alpha_{xx}\sim0.014$~$\mu$S for $\Delta=0.1$~eV in the HHP semiconductor, which is about one order of magnitude smaller than that in Cr$_2$O$_3$.
Given the fact that electric field--induced magnetization was measured in Cr$_2$O$_3$~\cite{Fiebig2005}, we believe that the magnetoelectric effect in the HHP semiconductor could be within the measurable window. 
Moreover, as already mentioned above, the band gap dependence in Fig.~\ref{fig:K_e12}(a) indicates that some semiconductors with a narrower gap could show a much larger magnetoelectric effect that is comparable to that in multiferroic materials. 
Finally, we note that the magnetoelectric effect presented in our work is qualitatively distinct from that in multiferroic materials: the former occurs on noninteracting systems and could appear in a frozen lattice, whereas the latter occurs on strongly-correlated systems and generally involves lattice distortions.

\section{Summary}\label{sec:summary}
In this work, we have theoretically demonstrated the magnetoelectric effect arising in heterostructures of a generic band insulator and a ferromagnet.
The observed magnetoelectric effect solely relies on the interband contributions between the valence and conduction bands, and hence, it is dissipationless, in contrast to the Edelstein effect in metals. 
By carefully studying the parameter dependences, we revealed that our scenario applies to generic situations: the only ingredients are the Rashba spin-orbit coupling inherent in the heterostructure and the exchange coupling from the proximity to the ferromagnet.
We also showed that the magnetoelectric signal is drastically enhanced around the resonant frequency corresponding to the band gap.
Bringing up a HHP semiconductor as a candidate material, we uncovered a considerable magnitude of the magnetoelectric effect.
Since the magnetoelectric mechanism in our results is quite general and readily expected in a wide range of band insulators in a heterostructure with a ferromagnet, our results could potentially offer a promising method for dissipationless electric generation and control of magnetization.

\begin{acknowledgments}
K.N.O. thanks N. Nagaosa and S. Murakami for illuminating discussions.
K.N.O. is supported by the Japan Society for the Promotion of Science through a research fellowship for young scientists.
This research was supported by JST CREST (JPMJCR18T2).
\end{acknowledgments}

\setcounter{equation}{0}

\appendix*
\section*{Appendix A: Tight-binding model for a generic band insulator}\label{sec:tb}
Here we deduce the tight-binding model in Eq.~(\ref{eq:toymodel}), which describes a generic two-dimensional band insulator in a heterostructure with a ferromagnet.
As stated in Sec.~\ref{sec:toymodel}, we assume two spherical orbitals named $1$ and $2$.
Here the Rashba spin-orbit coupling is effectively introduced through the electron hopping under the inversion-asymmetric potential.
We note that in real materials including HHP semiconductors introduced in Sec.~\ref{sec:HHPmodel}, the Rashba spin-orbit coupling is normally generated by combination of atomic spin-orbit coupling and hopping between orbitals with different parities, which is activated by the inversion-asymmetric potential.

First, we construct the electron hopping term in the tight-binding model.
The single-particle real-space Hamiltonian is given by
\begin{equation} 
H({\mathbf r})=\left[\frac{{\mathbf p}^2}{2m_{\rm e}}+V({\mathbf r})\right]I+\lambda{\bm\sigma}\cdot({\mathbf p}\times{\mathbf z}).
\label{eq:realspace}
\end{equation}
Here ${\mathbf p}=\,{}^{\rm t}\!(p_x, p_y)=-{\rm i}\hbar\,{}^{\rm t}\!(\partial_x, \partial_y)$ is the electron momentum, and $m_{\rm e}$ is the mass of an electron.
$V(\mathbf{r})$ is a periodic potential with $V(\mathbf{r})=V(\mathbf{r}+\mathbf{R})$, where $\mathbf{R}=n_x\mathbf{a}_x+n_y\mathbf{a}_y$ represents lattice vectors ($n_x, n_y$: integers; ${\mathbf a}_x$, ${\mathbf a}_y$: lattice vectors). 
The second term in Eq.~(\ref{eq:realspace}) represents the Rashba spin-orbit coupling arising from the electron hopping under the potential gradient along the $z$ axis, where $\lambda$ denotes the spin-orbit coupling strengh, and ${\mathbf z}$ is the unit vector along the $z$ axis. 
Based on Eq.~(\ref{eq:realspace}), the tight-binding Hamiltonian describing the electron hopping term is constructed on a square lattice as 
\begin{equation} 
\mathcal{H}_{\rm hop}=\sum_{ll'}\sum_{\rho\rho'}\sum_{ss'}t^{\rho s,\rho's'}_{ll'}c_{l\rho s}^\dagger c_{l'\rho' s'},
\label{eq:hhop}
\end{equation}
in which $c_{l\rho s}$ $(c^\dagger_{l\rho s})$ is the electron annihilation (creation) operator at site $l$ with orbital $\rho$ ($\rho=1$ and $2$) and spin $s$ ($s=\uparrow$ and $\downarrow$), and the transfer integrals $t^{\rho s,\rho's'}_{ll'}$ are written down as
\begin{equation} 
t^{\rho s,\rho's'}_{ll'}=\int d{\mathbf r}\phi^*_\rho({\mathbf r}-{\mathbf R}_l)H_{ss'}({\mathbf r})\phi_{\rho'}({\mathbf r}-{\mathbf R}_{l'}).
\label{eq:transfer_integrals}
\end{equation}
Here $\phi_\rho({\mathbf r})$ stands for the localized wave function of orbital $\rho$.
We take the summation $\sum_{ll'}$ over the nearest-neighbor (NN) sites as well as the identical sites with $l=l'$.

We analyzed the transfer integrals in Eq.~(\ref{eq:transfer_integrals}) based on the $C_{4v}$ symmetry of the wave function $\phi_\rho({\mathbf r})$ and periodic potential $V({\mathbf r})$.
The transfer integrals for the NN sites, denoted as $t^{\rho s,\rho's'}_{\pm{\mathbf a}_{x(y)}}=t^{\rho s,\rho's'}_{ll'}$ for ${\mathbf R}_{l'}-{\mathbf R}_l=\pm{\mathbf a}_{x(y)}$, are summarized as
\begin{align} 
t^{\rho s,\rho's'}_{\pm{\mathbf a}_x}&=\left(
    \begin{array}{cccc}
      t_1&\pm\lambda_1&t_{12}&\pm\lambda_{12}\\
      \mp\lambda_1&t_1&\mp \lambda_{12}&t_{12}\\
      t^*_{12}&\pm\lambda^*_{12}&t_2&\pm\lambda_2\\
      \mp\lambda^*_{12}&t^*_{12}&\mp\lambda_2&t_2\\
    \end{array}
  \right)\\
t^{\rho s,\rho's'}_{\pm{\mathbf a}_y}&=\left(
    \begin{array}{cccc}
      t_1&\mp {\rm i}\lambda_1&t_{12}&\mp {\rm i}\lambda_{12}\\
      \mp {\rm i}\lambda_1&t_1&\mp {\rm i}\lambda_{12}&t_{12}\\
      t^*_{12}&\mp {\rm i}\lambda^*_{12}&t_2&\mp {\rm i}\lambda_2\\
      \mp {\rm i}\lambda^*_{12}&t^*_{12}&\mp {\rm i}\lambda_2&t_2\\
    \end{array}
  \right), 
\end{align}
in which $t_{\rho\rho'}$ and $\lambda_{\rho\rho'}$ are described as
\begin{align} 
t_{\rho\rho'}&=\int d{\mathbf r}\phi^*_\rho({\mathbf r})\left[\frac{{\mathbf p}^2}{2m_{\rm e}}+V({\mathbf r})\right]\phi_{\rho'}({\mathbf r}-\mathbf{a}_x)\\
\lambda_{\rho\rho'}&={\rm i}\lambda\int d{\mathbf r}\phi^*_\rho({\mathbf r})p_x\phi_{\rho'}({\mathbf r}-\mathbf{a}_x),
\end{align}
and $t_\rho$ and $\lambda_\rho$ are defined as 
\begin{align} 
t_\rho&\equiv t_{\rho\rho}\\
\lambda_\rho&\equiv \lambda_{\rho\rho}.
\end{align}
Since $t_{\rho\rho'}=(t_{\rho'\rho})^*$ and $\lambda_{\rho\rho'}=(\lambda_{\rho'\rho})^*$ hold, $t_\rho$ and $\lambda_\rho$ are real values.
In the present work, we set a different sign between $t_1$ and $t_2$ to make a direct-gap semiconductor.
For simplicity, we set real values in $t_{12}$ and $\alpha_{12}$.

Likewise, the onsite transfer integrals, denoted as $t^{\rho s,\rho's'}_0=t^{\rho s,\rho's'}_{ll}$, are given by
\begin{equation} 
t^{\rho s,\rho's'}_0=\left(
    \begin{array}{cccc}
      \epsilon_1&0&0&0\\
      0&\epsilon_1&0&0\\
      0&0&\epsilon_2&0\\
      0&0&0&\epsilon_2\\
    \end{array}
  \right),
\end{equation}
where the onsite energies $\epsilon_\rho$ are described as 
\begin{equation} 
\epsilon_\rho=\int d{\mathbf r}\phi^*_\rho({\mathbf r})\left[\frac{{\mathbf p}^2}{2m_{\rm e}}+V({\mathbf r})\right]\phi_{\rho}({\mathbf r}).
\end{equation}
We set an energy offset as $\epsilon_1=-\frac{\epsilon_{12}}{2}$ and $\epsilon_2=\frac{\epsilon_{12}}{2}$.

Meanwhile, the exchange-coupling term in the tight-binding Hamiltonian is given by 
\begin{equation} 
\mathcal{H}_{\rm ex}=J_{\rm ex}\sum_l\sum_\rho\sum_{ss'}c_{l\rho s}^\dagger(M_x\sigma_x^{ss'}+M_z\sigma_z^{ss'})c_{l\rho s'}.
\label{eq:hex}
\end{equation}
Thus, via the Fourier transformation on the sum of Eqs.~(\ref{eq:hhop}) and (\ref{eq:hex}), we arrive at the momentum-space Hamiltonian shown in Eq.~(\ref{eq:toymodel}).

\section*{Appendix B: $\lambda_2$ dependence of the magnetoelectric effect}\label{sec:l2}
\begin{figure}[!htb]
\begin{center}
\includegraphics[width=0.8\columnwidth,clip]{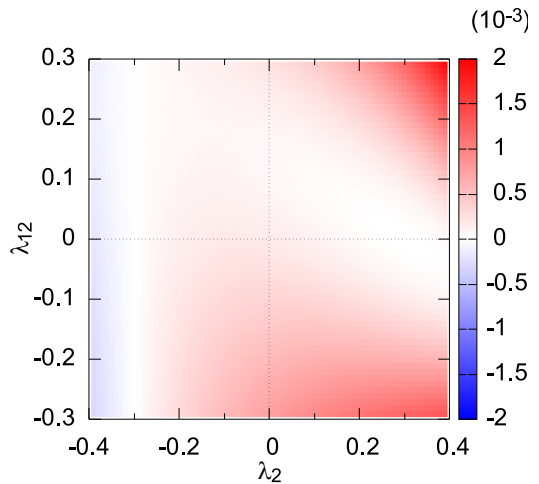}
\end{center}
\caption{$\lambda_2$-$\lambda_{12}$ dependence of $K_{xx}$. We set $J_{\rm ex}=0.1$ and ${\mathbf M}=(0, 0, 1)$ and use the same model parameters as in Fig.~\ref{fig:K_Jz}.}
\label{fig:K_l2}
\end{figure}

In this appendix, we study the dependence of the magnetoelectric effect on the intraband Rashba spin-orbit coupling $\lambda_2$ when the magnetic moment ${\mathbf M}$ points along the $z$ axis.
Figure~\ref{fig:K_l2} displays the $\lambda_2$-$\lambda_{12}$ dependence of $K_{xx}$ for $J_{\rm ex}=0.1$.
We set $t_1=1$, $\lambda_1=0.3$, $t_2=-1$, $\epsilon_{12}=8$, and $t_{12}=-0.2$ as in Fig.~\ref{fig:K_Jz}.
When $\lambda_2=0.3(=\lambda_1)$, $K_{xx}$ disappears for $\lambda_{12}=0$, which is also shown in Fig.~\ref{fig:K_e12}(b).
On the other hand, when $\lambda_2\neq0.3(=\lambda_1)$, $K_{xx}$ generally survives even for $\lambda_{12}=0$, except the case with $\lambda_2=-0.3(=-\lambda_1)$.

When $\lambda_2=-0.3$, Fig.~\ref{fig:K_l2} shows that $K_{xx}$ vanishes irrespective of $\lambda_{12}$.
This is guaranteed by particle-hole symmetry.
When $t_2=-t_1$ and $\lambda_2=-\lambda_1$ hold, the system is characterized with the particle-hole symmetry as $\Xi H(\mathbf{k})\Xi^{-1}=-H(-\mathbf{k})$, where the particle-hole operator $\Xi$ is represented as
\begin{align}
\Xi=\left(
    \begin{array}{cc}
      0&-{\rm i}\sigma_y\\
      {\rm i}\sigma_y&0
    \end{array}
    \right)K.
\end{align}
When the particle-hole symmetry is preserved, $K_{xx}$ in Eq.~(\ref{eq:K}) vanishes due to the cancellation between the contributions from ${\mathbf k}$ and $-{\mathbf k}$.

\bibliography{citation} 

\end{document}